\documentclass[aps,prc,twocolumn,groupedaddress,showpacs,nofootinbib,fleqn,floatfix,superscriptaddress]{revtex4}
\usepackage{epsfig,amsmath,amssymb,graphicx,tabularx,dcolumn,bm,longtable}
\usepackage[usenames,dvipsnames]{color}
\definecolor{orange}{rgb}{0.9,0.45,0}

\begin{document}
\bibliographystyle{apsrev}

\title{Multipole strength function of deformed superfluid nuclei made easy}

\author{M.~Stoitsov}
\affiliation{Department of Physics \& Astronomy, University of Tennessee, Knoxville, Tennessee 37996, USA}
\affiliation{Physics Division,  Oak Ridge National Laboratory, P.O. Box 2008, Oak Ridge, Tennessee 37831, USA}

\author{M.~Kortelainen}
\affiliation{Department of Physics \& Astronomy, University of Tennessee, Knoxville, Tennessee 37996, USA}
\affiliation{Physics Division,  Oak Ridge National Laboratory, P.O. Box 2008, Oak Ridge, Tennessee 37831, USA}

\author{T.~Nakatsukasa}
\affiliation{RIKEN Nishina Center, Wako-shi, 351-0198, Japan}
\affiliation{Center for Computational Sciences, University of Tsukuba, Tsukuba, 305-8571, Japan}

\author{C.~Losa}
\affiliation{International School for Advanced Studies, SISSA, Via Bonomea 265 34136 Trieste, Italy}

\author{W.~Nazarewicz}
\affiliation{Department of Physics \& Astronomy, University of Tennessee, Knoxville, Tennessee 37996, USA}
\affiliation{Physics Division,  Oak Ridge National Laboratory, P.O. Box 2008, Oak Ridge, Tennessee 37831, USA}
\affiliation{Institute of Theoretical Physics, Warsaw University, ul. Ho\.za 69, 00-681 Warsaw, Poland}

%\date{\today}

\begin{abstract}

We present an efficient method for calculating strength functions
using the finite amplitude method (FAM)
for deformed superfluid heavy nuclei within the framework of the nuclear density functional theory. We demonstrate
that FAM reproduces strength functions obtained with the fully self-consistent quasi-particle random-phase approximation  (QRPA) 
at a fraction  of computational cost. As a demonstration, we compute the
isoscalar and isovector monopole strength  for  strongly deformed configurations in
$^{240}$Pu by considering huge quasi-particle QRPA spaces.
Our approach  to FAM,  based on Broyden's iterative procedure,
opens the possibility for large-scale calculations of strength distributions in well-bound and weakly bound nuclei across the nuclear landscape.

\end{abstract}


\maketitle

%\section{Introduction}\label{sec0}
{\it Introduction.}--
One of the major challenges in the  many-body problem is the microscopic description of the  collective motion involving hundreds of strongly interacting particles. Here, of particular interest is the response of the system to
a time-dependent external field. In the nuclear case, in the small-amplitude limit of nearly harmonic oscillations about equilibrium,
the phenomena of interest include the variety of vibrational modes, and characteristic distribution of electromagnetic, particle, and  beta-decay strength \cite{(Boh75),(ring)}.

Most nuclei exhibit strong nucleonic pairing that profoundly impacts
their collective motion \cite{(Row70),[Bri05]}.
When moving away from the stability line, another factor affecting nuclear correlations, and dynamics  is the presence of a low-lying particle continuum 
which provides a vast configuration space for nucleonic  excitations. 
Therefore, to understand the variety of nuclear modes throughout the nuclear chart, 
a consistent  treatment of many-body correlations and continuum is required \cite{[Dob07b]}.

This study is devoted to the multipole strength in superfluid deformed heavy nuclei. The theoretical method is the quasi-particle
random-phase approximation (QRPA) applied to the self-consistent  configuration obtained by means of the nuclear density 
functional theory (DFT) \cite{[Ben03]}. 
QRPA represents a small amplitude limit to the  time-dependent superfluid DFT  method.
In the absence of pairing, QRPA reduces to the usual Random Phase Approximation (RPA) built atop the Hartree-Fock (HF) equilibrium.

In the electronic DFT,
the RPA contribution to electron correlation energies has emerged \cite{[rpadft],[Furche08]}  as
an important building block of accurate density functional treatments of molecules and solids as it combines a number
of attractive features: it includes  the long-range dispersion \cite{[lld]} as opposed to semi-local functionals;  it is non-perturbative and can
be applied to small or zero gap problems, such as metals~\cite{[metals]} or dissociating H$_2$~\cite{[dissociating]}; it is nearly exact in the high-density or low-coupling limit, and it is parameter-free; finally  it is intimately related to the microscopic Coupled Cluster doubles  theory 
\cite{[Freeman76],[Scuseria08],[Tretiak]}.
In the nuclear DFT, the applicability of (Q)RPA to correlation energies is more limited \cite{[Don99],[Shi00],[Hag00]} as many nuclei have transitional character, i.e., they are close to the 
critical point for the symmetry-breaking where the second-order expansion in density fluctuations breaks down~\cite{[Naz93]}. In spite of its
drawbacks, because of its deep connection to DFT, QRPA is still the tool of choice when  it comes to either spherical or well-deformed nuclei.
In addition to
numerous applications to small-amplitude collective motion, 
QRPA for deformed nuclei may be utilized in the calculation of the
collective mass for the large-amplitude dynamics \cite{[Hin10]}.

The advantage of QRPA+DFT  is that it properly takes into account  self-consistent couplings giving rise to the variety of  symmetry-breaking phenomena and, when done properly, includes the effects due to the continuum coupling. Due to the complexity of the problem, the QRPA framework being capable of a fully self-consistent description of non-spherical systems has eluded us until very recently
\cite{[Ter10],[Losa10],[Terasaki:11],[Peru.ea:11],[Martini:11]}. 

A major obstacle preventing  the widespread use of QRPA  has been its high computational cost. In chemistry, this factor has limited
 applications of this method considerably~\cite{[Furche08]}. In nuclear physics, most  fully self-consistent QRPA applications have been  limited to
spherical nuclei (see, e.g., \cite{[jun05]} and references therein).  In spite of advanced computational resources available, it is only very recently that deformed QRPA calculations atop the Hartree-Fock-Bogoliubov (HFB) equilibrium   have been carried out
\cite{[Ter10],[Losa10],[Terasaki:11],[Peru.ea:11],[Martini:11]}.

The primary challenge in the conventional matrix formulation of (Q)RPA is computation  and storage 
of huge matrices of the residual interaction.
The recent breakthrough came from the realization that both bottlenecks  can be avoided by taking advantage of self-consistent DFT solutions and directly employing  the linear response theory to them (see literature quoted in Refs.~\cite{[Tretiak],[Nak07],[Toi10]}). Indeed, in both the
finite-amplitude method (FAM) \cite{[Nak07]}
and the iterative non-Hermitian Arnoldi diagonalization technique \cite{[Toi10]},
 the (Q)RPA  matrix equations are recast into the set of
self-consistent equations with respect to (Q)RPA amplitudes,
which significantly reduces  computational effort.
The FAM  has been applied in the RPA variant to Skyrme energy density functionals (EDFs) to study
giant dipole resonances and low-lying pygmy dipole modes
 \cite{[Inakura09],[Inakura11]}. Recently, in its QRPA extension, FAM 
was used to study monopole resonances in a  spherical drip-line nucleus $^{174}$Sn \cite{[Avo11]}.
The iterative Arnoldi diagonalization was first used in the RPA variant
to electromagnetic strength functions in $^{132}$Sn \cite{[Toi10]}, and
the spherical QRPA extension has recently been reported  \cite{ArnoldiQRPA}.

While based on the same principle, the actual numerical implementations of FAM 
and iterative Arnoldi diagonalization differ.
In the applications of FAM, the (Q)RPA amplitudes
are iterated at desired energies. The  Arnoldi algorithm  is a  Krylov subspace iterative method for extracting a {\it partial} eigenspectrum, i.e., it computes a discrete set of states that approximate true eigenvectors.
(For another nuclear application, an iterative Lanczos-based power iteration algorithm for solving the RPA equations, see  Ref.~\cite{[Johnson]}.)

The current implementation of FAM has so far been done in the
coordinate-space representation that
 requires
a large number of iterations (in some cases more than
500-1,000 \cite{[Nak07],[Inakura09]}) to obtain self-consistent amplitudes.
Here, we propose a fast and efficient method for solving the FAM-QRPA
equations in the harmonic oscillator (HO) basis
using the Broyden iterative scheme \cite{[broyden],[gbroyden]}  previously  adopted to HFB equations of  nuclear DFT \cite{[hfbbroyden]}.
We study the performance of the method and compare it with the standard QRPA
diagonalization method. We demonstrate that FAM-QRPA solutions can be  obtained with no more than 40 iterations  
at modest memory requirements of about half a gigabyte
for large model spaces corresponding to extreme cases of fission isomers in the actinides.

%\section{FINITE AMPLITUDE METHOD}\label{secFAM}
{\it Finite Amplitude Method.}--
The basic formulation of the FAM-RPA is presented in Ref. \cite{[Nak07]},
and that for the FAM-QRPA is given in Ref. \cite{[Avo11]}. The implementation of a traditional matrix formulation  QRPA method (MQRPA) used in this work follows that of Ref.~\cite{[Losa10]}.
In this section, we recapitulate the method and define all necessary quantities.

%\subsection{Matrix Formulation of QRPA}
The variation of the total DFT energy defined through an  energy density 
${\cal E}(\rho,\kappa, \kappa^*)$, with respect to  the particle and pairing  
density matrices,
$\rho=V^{\ast }V^{T}$ and $\kappa=V^{\ast }U^{T}$, results in the HFB equations
\begin{equation}  \label{hfb}
\left(
\begin{array}{cc}
h -\lambda & \Delta \\
-\Delta^{\ast } & -h^{\ast }+\lambda
\end{array}
\right)
\left(
\begin{array}{c}
U_\mu \\
V_\mu
\end{array}
\right) =
E_\mu
\left(
\begin{array}{c}
U_\mu \\
V_\mu
\end{array}
\right) ,
\end{equation}
where
\begin{equation}
 h_{kl}[\rho,\kappa,\kappa^*]
=\frac{\partial {\cal E}[\rho,\kappa,\kappa^*]}{\partial \rho_{lk}},~
 \Delta_{kl}[\rho,\kappa]
=\frac{\partial {\cal E}[\rho,\kappa,\kappa^*]}{\partial \kappa^{*}_{kl}},
\label{hfbeqhd}
\end{equation}
$E_\mu$ are the quasi-particle energies, ($U_\mu,V_\mu$) are the two-component HFB quasi-particle vectors, and
the chemical potential $\lambda $ is introduced to conserve the 
 average particle number.

The QRPA equations for the mode  amplitudes ($X_{\mu \nu}$, $Y_{\mu \nu}$) and  excitation energies $\omega$
can be written in a matrix form as:
\begin{equation}
\left(
\begin{array}{cc}
A & B \\
B^* & A^*
\end{array}
\right) \left(
\begin{array}{c}
X \\
Y
\end{array}
\right) =\omega \left(
\begin{array}{c}
~ X \\
- Y
\end{array}
\right)   \label{qrpa}
\end{equation}
with  matrices A and B coming from  second variational derivatives of ${\cal E}[\rho,\kappa,\kappa^*]$ with respect to $\rho$ and $\kappa$. In MQRPA, Eqs.  (\ref{qrpa}) are solved  by means of the explicit diagonalization,  and the strength function corresponding to the 
one-body operator $\hat F$ is subsequently computed.
In our results, strength functions calculated with MQRPA are
smeared with a Lorentzian-averaging function having  a width $\Gamma=2 \gamma$.
Such an averaging can be associated with complex QRPA frequencies
$\omega + i\gamma$ that are introduced in the context of strength function technique  with schematic  interactions \cite{(Boh75),[Malov76],[Malov77],[Kvasil98]}.
In fact,  strength functions obtained in such a way do not require knowledge of individual RPA eigenvalues; the summation over the RPA spectrum is replaced by integration over energy (see also Refs.~\cite{[Don99],[Shi00]}).

%\subsection{FAM-QRPA Formulation}

Following the earlier applications of FAM \cite{[Nak07],[Inakura09],[Inakura11],[Avo11]},
we solve the QRPA problem in the presence of a one-body external perturbation $\hat F$ of a given frequency $\omega$. In this case, 
Eq.~(\ref{qrpa}) can be rewritten in an alternative way \cite{[Avo11]}:
\begin{equation}
\begin{array}{c}
\left(E_\mu + E_\nu -\omega\right) X_{\mu \nu} +\delta H^{20}_{\mu \nu}(\omega) =F^{20}_{\mu \nu} ,  \\[4pt]
\left(E_\mu + E_\nu + \omega \right) Y_{\mu \nu} +\delta H^{02}_{\mu \nu}(\omega) =F^{02}_{\mu \nu} ,
\label{FAM}
\end{array}
\end{equation}
where the complex antisymmetric matrices 
\begin{equation}
\begin{array}{rll}
\displaystyle \delta H^{20}(\omega) &
= \displaystyle U^\dagger \delta h(\omega) V^* & -~ V^\dagger \delta h(\omega)^T U^*      \\
 &- \displaystyle   V^\dagger \overline{\delta \Delta}^{*}(\omega) V^* & +~ U^\dagger \delta \Delta(\omega) U^* , \\[4pt]
\delta H^{02}(\omega) &
= \displaystyle U^T \delta h(\omega)^T V & -~ V^T \delta h(\omega) U    \\
 &\displaystyle  -V^T \delta\Delta(\omega) V & +~ U^T \overline{\delta \Delta}(\omega)^* U,
\label{H2}
\end{array}
\end{equation}
are defined in terms of  the non-Hermitian   variations
\begin{equation}
\begin{array}{rl}
\displaystyle \delta h(\omega) &= \displaystyle (h[\rho_\eta,\kappa_\eta,\bar{\kappa}_\eta]-h[\rho,\kappa,\kappa^*])/\eta  , \\[4pt]
\displaystyle \delta \Delta (\omega) &= \displaystyle (\Delta[\rho_\eta,\kappa_\eta]-\Delta[\rho,\kappa])/\eta , \\[4pt]
\displaystyle \overline{\delta  \Delta} \displaystyle (\omega) & = (\Delta[\bar{\rho}_\eta,\bar{\kappa}_\eta]-\Delta[\rho,\kappa])/\eta,
\label{hd}
\end{array}
\end{equation}
where $\eta$ is a small parameter to numerically expand densities up to
the first order.
The non-Hermitian density matrices in (\ref{hd}) are:
\begin{equation}
\begin{array}{rl}
\displaystyle \rho_\eta =&\displaystyle ~ \left(V + \eta U^* X^* \right)^*  \left(V + \eta U^* Y\right)^T  ,  \\[2pt]
\displaystyle \kappa_\eta = & \displaystyle - \left(U + \eta V^* Y\right) \left(V + \eta U^* X^* \right)^\dagger ,  \\[2pt]
\displaystyle \bar{\rho}_\eta =& \displaystyle ~ \left(V + \eta U^* Y \right)^*  \left(V + \eta U^* X^*\right)^T  ,  \\[2pt]
\displaystyle \bar{\kappa}_\eta = &\displaystyle - \left(U + \eta V^* X^*\right) \left(V + \eta U^* Y \right)^\dagger.
\label{rketa}
\end{array}
\end{equation}
We note that the above density matrices depend on the external field $\hat F$ through the QRPA vectors ($X,Y$).

The FAM-QRPA equations~(\ref{FAM}) can be formally solved with respect to $X_{\mu\nu}$,   $Y_{\mu\nu}$:
\begin{equation}
%\begin{array}{rl}
%\displaystyle X_{\mu\nu} &= \displaystyle - \frac{\delta H^{20}_{\mu\nu}(\omega)-F^{20}_{\mu\nu} }{E_\mu+E_\nu-\omega}  ,  \\[10pt]
%\displaystyle Y_{\mu\nu} &= \displaystyle - \frac{\delta H^{02}_{\mu\nu}(\omega)-F^{02}_{\mu\nu} }{E_\mu+E_\nu+\omega} .
X_{\mu\nu} = - \frac{\delta H^{20}_{\mu\nu}(\omega)-F^{20}_{\mu\nu} }{E_\mu+E_\nu-\omega} , ~~~
Y_{\mu\nu} = - \frac{\delta H^{02}_{\mu\nu}(\omega)-F^{02}_{\mu\nu} }{E_\mu+E_\nu+\omega} .
\label{xy}
%\end{array}
\end{equation}
Since the matrices $\delta H^{20}(\omega)$ and  $\delta H^{02} (\omega)$
linearly depend on $X_{\mu\nu}$ and  $Y_{\mu\nu}$,
a self-consistent  iterative scheme needs to be adopted to find the QRPA amplitudes.
In essence, 
FAM replaces the calculation and  diagonalization of the large QRPA matrices $A$,  $B$  with a much
simpler procedure of   calculating  $\delta H^{20} (\omega)$ and  $\delta {H}^{02} (\omega)$, 
and solving Eqs.~(\ref{FAM}) at desired values of $\omega$. To guarantee that
the FAM-QRPA solution carries a finite strength function for every value  of  $\omega$,
we take $\omega \rightarrow \omega+ i \gamma$ with a small 
imaginary part $\gamma$. As we shall see later, such a choice corresponds to a Lorentzian smearing of $\Gamma=2 \gamma$, except for the vicinity of $\omega=0$ \cite{(Boh75)}.
It is worth noting that the FAM  implementation of QRPA  is straightforward and EDF-independent, as the same HFB procedures that define the fields $h,\Delta$ in terms of particle and pairing densities  are also used in FAM-QRPA calculations.

%\subsection{Iterative procedure used in FAM}

Ideally, a self-consistent iterative FAM-QRPA  procedure should converge rapidly and the result should not depend  on  $\eta$ if its value is small enough.
In practice,  a direct iteration  of (\ref{xy}) diverges in most cases,
especially when  $\gamma$ is small and/or when $\omega$ is close to the true QRPA root. 
Indeed, when  $\gamma\to 0$, the left-hand side of Eqs.~(\ref{FAM}) becomes singular; hence, instabilities are expected around QRPA roots.
To guarantee numerical 
stability, one has to resort to a  procedure which `mixes' the solutions from previous and next iterations. To this end,
the conjugate gradient method and its derivatives were utilized in  coordinate-space applications
\cite{[Nak07],[Inakura09],[Inakura11],[Avo11]}.
In this work, based on the HO expansion technique,
the modified Broyden's procedure \cite{[gbroyden],[hfbbroyden]}
has been adopted and turned out to yield stable results while providing excellent computational performance. 
For a system of linear equations (\ref{FAM}), Broyden iterations exhibit  the $Q$-superlinear convergence   \cite{[Kelley95]}.

In the Arnoldi (or Lanczos) diagonalization scheme, QRPA iterations
start from a pivot vector that has
its elements set to the matrix elements of $\hat F$. This guarantees that  
all odd-power energy-weighed
sum rules (EWSR) are conserved. The number of states found is equal to the number of iterations, and these states 
(usually different from QRPA modes) are used to construct the strength function.
On the other hand, by solving the linear response equation (\ref{FAM})
with a fixed $\omega$ is that one obtains solutions at required  energies. In this way the individual 
QRPA states can be further  accessed, if needed.

%\section{RESULTS}\label{secresults}
{\it Results.}--
To benchmark FAM-QRPA for deformed nuclei and 
check its performance, we carried out FAM and MQRPA  \cite{[Losa10]}  calculations using the SLy4 EDF \cite{sly4}
and a contact volume pairing with a 60 MeV cutoff with respect to the reference single-particle energies \cite{[Sto05]}. The pairing strength was chosen to reproduce the experimental  neutron pairing gap of $^{120}$Sn. 
All HFB calculations were performed with the DFT code HFBTHO 
\cite{[Sto05]} that solves the Skyrme-HFB equations in the HO basis, assuming axial and reflection symmetries.

As discussed in Ref.~\cite{[Losa10]},  MQRPA calculations are subject to two truncations. The first truncation pertains to the maximum rank  of the QRPA Hamiltonian matrix that, in our case,  should  not exceed 20,000. To this end, one neglects all canonical states  with single-particle energies  greater than some cutoff value. The 
second truncation is made by excluding those QRPA quasiparticle pairs that have occupation probabilities less than some small critical value
$v^2_{\rm crit}$ or larger than $1-v^2_{\rm crit}$. 

The benchmarking calculations have been carried out for the monopole isoscalar (IS)  and modified isovector  (IV) response operators:
\begin{equation}
f^{IS} =  \displaystyle  \frac{ e Z}{A} \sum_{i=1}^{A} r_i^2,~~~f^{IV} =  \displaystyle  \frac{ e Z}{A} \sum_{i=1}^{N} r_i^2 - \frac{ e N}{A} \sum_{i=1}^{Z} r_i^2.
\label{fisiv} 
\end{equation}
This choice makes coupling between IS and IV monopole vibrations small \cite{[Lip89]}.
At each value   $\omega$, the  imaginary part  has been set to $\gamma=0.5$ MeV, and the FAM strength function has 
been calculated according to  \cite{[Avo11]}
\begin{equation}
\displaystyle S( f, \omega ) =  \displaystyle - \frac{1}{\pi \alpha}  {\rm Im~Tr} \left[ f (U X V^T + V^* Y^T U^\dagger )\right] .
\label{aa}
\end{equation}
Here, the external field in Eq. (\ref{FAM}) is given by
$F=\alpha f$, where $\alpha$ is a parameter with dimension
$[\alpha]=[F][f]^{-1}$.
Since for very small values of $\eta$ the QRPA amplitudes $X$ and $Y$ are proportional to $\alpha$,
$S(f,\omega)$ is independent of $\alpha$.
Using a complex frequency  $\omega+i\gamma$,
the resulting  $S(f,\omega)$ possesses the
crossing symmetry $S(f,\omega)=-S(f,-\omega)$ \cite{(Boh75)};
hence $S(f,\omega=0)=0$ is guaranteed.
The strength function obtained in  MQRPA  has been computed 
by averaging  QRPA diagonalization results with 
$\Gamma=2 \gamma =1$ MeV. In the fully self-consistent framework,
the MQRPA and FAM-QRPA results should be identical.

\begin{figure}[htb]
\includegraphics[width=1.0\columnwidth]{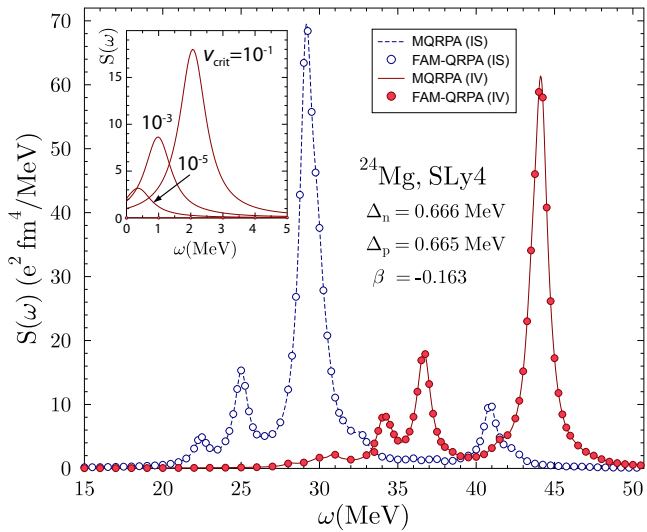}
\caption{(color online)  The  isoscalar (blue, dashed line)  and isovector (red, solid line) monopole strength  function in oblate-deformed and paired minimum of $^{24}$Mg obtained in MQRPA  within the full $N_{\rm sh}=5$ HO space ($v_{\rm crit}$=0) and  FAM-QRPA (circles). The inset shows the presence of 0$^+$ spurious mode at low-energy 
at three different values of $v_{\rm crit}$ in MQRPA.}
\label{fig1}
\end{figure}
The  MQRPA and FAM-QRPA $J^\pi=0^+$ strength functions are compared in Fig.~\ref{fig1} for an oblate  minimum in  $^{24}$Mg. (In the prolate ground state of this nucleus, paring correlations vanish.)  To include the whole  available space of canonical wave functions in MQRPA,
the results shown in Fig.~\ref{fig1} were obtained using a relatively small
single-particle basis corresponding to $N_{\rm sh}=5$ HO shells. 
It is clearly seen  that  both methods yield practically identical results.
Increasing the number of basis states rapidly increases the scale of the MQRPA  scheme.
For example, with 20 HO shells and $v_{\rm crit}=10^{-4}$, 
QRPA matrices reach  dimension 32,039, requiring 16.4 GB memory. Lowering the canonical cutoff to   $v_{\rm crit}=10^{-20}$ results 
in the matrix rank   211,159, or 713.41 GB of memory.  
In contrast, the memory required by FAM-QRPA, using the full space of 20 HO shells and without any 
truncations on a QRPA level, is a modest 0.572 GB. 

The accuracy of any QRPA implementation can be assessed by its ability to handle the spurious (zero-energy) modes. In  general, QRPA solutions should be properly orthogonalized against  spurious  modes by means of a well-established procedure \cite{[Nak07],[Toi10]}. For the monopole case presented in Fig.~\ref{fig1}, the huge configuration space of FAM-QRPA seems to be sufficient to remove the 0$^+$ spurious modes associated with particle-number breaking  almost exactly. This is indeed seen in the inset of Fig.~\ref{fig1}
which displays the low-energy IV monopole strength in MQRPA for several 
values of $v_{\rm crit}$. The MQRPA response corresponding to
$v_{\rm crit}$=0 (full HO space) shows the single low-energy peak 
carrying the strength of about 2$\cdot$10$^{-6}$\,$e^2$fm$^4$/MeV, and
the low-energy FAM-QRPA strength is of similar magnitude.

\begin{figure}[htb]
\includegraphics[width=0.80\columnwidth]{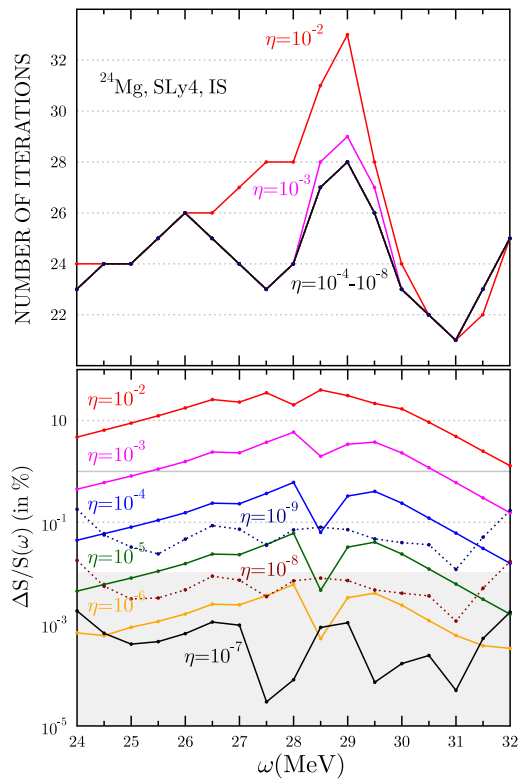}
\caption{(color online)  The performance of the FAM-QRPA algorithm applied to  the case of Fig.~\ref{fig1} for different values of $\eta$  in the frequency range of 24$<$$\omega$$<$32\,MeV. Top: number of  Broyden iterations. Bottom:  relative accuracy.}
\label{fig2}
\end{figure}

\begin{figure}[htb]
\includegraphics[width=0.88\columnwidth]{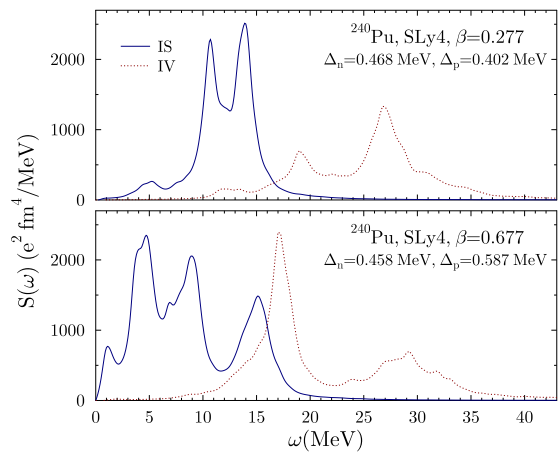}
\caption{(color online)  IS (solid line) and IV (dotted line) monopole strength in the  deformed ground state of  $^{240}$Pu and its superdeformed fission isomer in FAM-QRPA with $N_{\rm sh}$=20.}
\label{fig4}
\end{figure}

Figure~\ref{fig2} demonstrates  that the FAM results practically do not depend on the choice of the parameter $\eta$ entering the numerical derivatives in Eq. (\ref{hd}) for quite a large range of values of $\eta$ from $10^{-6}$ to  $10^{-8}$. Usually, the FAM solution is  reached fairly quickly,  within 10-40 iterations assuming  that the maximum difference between collective amplitudes in two consecutive iterations is less that $10^{-6}$.

In order to demonstrate the ability of FAM-QRPA to treat heavy, deformed, and superfluid  nuclei, Fig.~\ref{fig4} shows the monopole strength distributions for the ground state and fission isomer in $^{240}$Pu obtained with a large configuration space of 
$N_{\rm sh}=20$. As it is well known \cite{[Garg80]}, due to its large deformation,
the IS monopole strength distribution splits into two components. 
Here, as well as  in other cases considered in this work, about  99$\%$  of the EWSR is exhausted 
when integrating up to $\omega$=50\,MeV.
Both examples nicely  illustrate the applicability of  FAM-QRPA  to the local QRPA approach used in the context of the 
large-amplitude collective motion \cite{[Hin10]}.
%%

%\section{CONCLUSIONS}\label{secconclusions}
{\it Conclusions.}--
In this work we applied the FAM-QRPA method to describe the multipole strength
in deformed and superfluid nuclei. The calculations have been presented for IS and IV monopole modes. 
We first benchmarked FAM-QRPA against the MQRPA approach  of Ref.~\cite{[Losa10]} and obtained excellent 
agreement. As compared  to  the standard diagonalization method, FAM-QRPA offers 
excellent  performance, both in terms of memory and speed.
Including all the fields (both time-even and time-odd) required by the fully self-consistent QRPA,  
the memory requirement for the FAM-QRPA module built on the top of the DFT solver HFBTHO does not exceed  572 MB. 
This enables us to handle 
axially-deformed  heavy nuclei  without imposing any truncation on the QRPA level. 
The self-consistency of FAM-QRPA, together  with very large pairing windows employed, results in a
practical decoupling of  the  0$^+$ spurious modes
associated with the particle-number symmetry breaking.

A new  efficient and robust  method for the iterative  solution of FAM-QRPA equations has been proposed. The method, based on the
Broyden mixing procedure already adopted in DFT solvers \cite{[hfbbroyden]},
offers a $Q$-superlinear convergence of FAM-QRPA equations.
The proposed  FAM implementation allows fast  calculations  of multipole strength for all axially deformed nuclei throughout 
the nuclear  chart. The algorithm is especially suited for
multiprocessor tasks since  the QRPA strength distribution  $S(\omega)$  converges with no more then 40 iterations regardless of $\omega$.
The implementation of the method to  higher multipolarity  modes is in progress.

%\begin{acknowledgments}
The authors thank J. Dobaczewski and K. Matsuyanagi for valuable discussions.
This work was supported in part
by the U.S. Department of Energy under Contract Nos.
DE-FG02-96ER40963 (University of Tennessee),
DE-FC02-09ER41583 (UNEDF SciDAC Collaboration), and DE-FG02-06ER41407 (JUSTIPEN), and by  KAKENHI of JSPS (No. 21340073 and No. 20105003).
%\end{acknowledgments}

%\bibliography{famqrpa_etal}

\begin{thebibliography}{43}
\expandafter\ifx\csname natexlab\endcsname\relax\def\natexlab#1{#1}\fi
\expandafter\ifx\csname bibnamefont\endcsname\relax
  \def\bibnamefont#1{#1}\fi
\expandafter\ifx\csname bibfnamefont\endcsname\relax
  \def\bibfnamefont#1{#1}\fi
\expandafter\ifx\csname citenamefont\endcsname\relax
  \def\citenamefont#1{#1}\fi
\expandafter\ifx\csname url\endcsname\relax
  \def\url#1{\texttt{#1}}\fi
\expandafter\ifx\csname urlprefix\endcsname\relax\def\urlprefix{URL }\fi
\providecommand{\bibinfo}[2]{#2}
\providecommand{\eprint}[2][]{\url{#2}}

\bibitem[{\citenamefont{{A. Bohr and B.R. Mottelson, {\it Nuclear Structure\/},
  Vol. I (W.A. Benjamin, New York, 1969); Vol. II (W.A. Benjamin, New York,
  1975)}}()}]{(Boh75)}
\bibinfo{author}{\bibnamefont{{A. Bohr and B.R. Mottelson, {\it Nuclear
  Structure\/}, Vol. I (W.A. Benjamin, New York, 1969); Vol. II (W.A. Benjamin,
  New York, 1975)}}}.

\bibitem[{\citenamefont{Ring and Schuck}(1980)}]{(ring)}
\bibinfo{author}{\bibfnamefont{P.}~\bibnamefont{Ring}} \bibnamefont{and}
  \bibinfo{author}{\bibfnamefont{P.}~\bibnamefont{Schuck}},
  \emph{\bibinfo{title}{The Nuclear Many-Body Problem}}
  (\bibinfo{publisher}{Springer Verlag}, \bibinfo{address}{New York},
  \bibinfo{year}{1980}).

\bibitem[{\citenamefont{Rowe}(1970)}]{(Row70)}
\bibinfo{author}{\bibfnamefont{D.}~\bibnamefont{Rowe}},
  \emph{\bibinfo{title}{Nuclear Collective Motion, Models and Theory}}
  (\bibinfo{publisher}{Mathuen, London}, \bibinfo{year}{1970}).

\bibitem[{\citenamefont{{D.M. Brink and R.A. Broglia, {\it Nuclear
  Superfluidity: Pairing In Finite Systems} (Cambridge Univ. Press, Cambridge,
  2005)}}()}]{[Bri05]}
\bibinfo{author}{\bibnamefont{{D.M. Brink and R.A. Broglia, {\it Nuclear
  Superfluidity: Pairing In Finite Systems} (Cambridge Univ. Press, Cambridge,
  2005)}}}.

\bibitem[{\citenamefont{{J. Dobaczewski  {\it et al.},
  Prog. Part. Nucl. Phys. {\bf 59}, 432 (2007)}}()}]{[Dob07b]}
\bibinfo{author}{\bibnamefont{{J. Dobaczewski {\it et
  al.}, Prog. Part. Nucl. Phys. {\bf 59}, 432 (2007)}}}.

\bibitem[{\citenamefont{{M. Bender, P.-H. Heenen, and P.-G. Reinhard, Rev. Mod.
  Phys. {\bf 75}, 121 (2003)}}()}]{[Ben03]}
\bibinfo{author}{\bibnamefont{{M. Bender, P.-H. Heenen, and P.-G. Reinhard,
  Rev. Mod. Phys. {\bf 75}, 121 (2003)}}}.

\bibitem[{\citenamefont{{D. C. Langreth and J.P. Perdew, Phys. Rev. B
  \textbf{15}, 2884 (1977).}}()}]{[rpadft]}
\bibinfo{author}{\bibnamefont{{D. C. Langreth and J.P. Perdew, Phys. Rev. B
  \textbf{15}, 2884 (1977).}}}

\bibitem[{\citenamefont{{F. Furche, J. Chem. Phys. \textbf{129}, 114105
  (2008)}}()}]{[Furche08]}
\bibinfo{author}{\bibnamefont{{F. Furche, J. Chem. Phys. \textbf{129}, 114105
  (2008)}}}.

\bibitem[{\citenamefont{{J. Dobson, in {\it Time-Dependent Density Functional
  Theory}, Vol. 706, p 443 (Springer, Berlin Heidelberg, 2006)}}()}]{[lld]}
\bibinfo{author}{\bibnamefont{{J. Dobson, in {\it Time-Dependent Density
  Functional Theory}, Vol. 706, p 443 (Springer, Berlin Heidelberg, 2006)}}}.

\bibitem[{\citenamefont{{J. Harl and G. Kresse, Phys. Rev. B \textbf{77},
  045136 (2008); Phys. Rev. Lett., \textbf{103}, 056401 (2009)}}()}]{[metals]}
\bibinfo{author}{\bibnamefont{{J. Harl and G. Kresse, Phys. Rev. B \textbf{77},
  045136 (2008); Phys. Rev. Lett., \textbf{103}, 056401 (2009)}}}.

\bibitem[{\citenamefont{{F. Weigend  {\it et al.},
  Chem. Phys. Lett., \textbf{294}, 143 (1998); M. Fuchs {\it et al.}, Chem. Phys., \textbf{122}, 094116
  (2005)}}()}]{[dissociating]}
\bibinfo{author}{\bibnamefont{{F. Weigend {\it et
  al.}, Chem. Phys. Lett., \textbf{294}, 143 (1998); M. Fuchs {\it et al.}, Chem. Phys., \textbf{122}, 094116 (2005)}}}.

\bibitem[{\citenamefont{{D.L. Freeman and M.J. Karplus, Chem. Phys.
  \textbf{64}, 2641 (1976)}}()}]{[Freeman76]}
\bibinfo{author}{\bibnamefont{{D.L. Freeman and M.J. Karplus, Chem. Phys.
  \textbf{64}, 2641 (1976)}}}.

\bibitem[{\citenamefont{{G.E. Scuseria, T.M. Henderson, and D.C. Sorensen, J.
  Chem. Phys. \textbf{129}, 231101 (2008)}}()}]{[Scuseria08]}
\bibinfo{author}{\bibnamefont{{G.E. Scuseria, T.M. Henderson, and D.C.
  Sorensen, J. Chem. Phys. \textbf{129}, 231101 (2008)}}}.

\bibitem[{\citenamefont{{S. Tretiak {\it et
  al.}, J. Chem. Phys. {\bf 130}, 054111 (2009)}}()}]{[Tretiak]}
\bibinfo{author}{\bibnamefont{{S. Tretiak {\it et al.}, J. Chem. Phys. {\bf 130}, 054111 (2009)}}}.

\bibitem[{\citenamefont{{F. D\"onau, D. Almehed, and R.G. Nazmitdinov, Phys.
  Rev. Lett. {\bf 83}, 280 (1999)}}()}]{[Don99]}
\bibinfo{author}{\bibnamefont{{F. D\"onau, D. Almehed, and R.G. Nazmitdinov,
  Phys. Rev. Lett. {\bf 83}, 280 (1999)}}}.

\bibitem[{\citenamefont{{Y.R. Shimizu, P. Donati, and R.A. Broglia, Phys. Rev.
  Lett. {\bf 85}, 2260 (2000)}}()}]{[Shi00]}
\bibinfo{author}{\bibnamefont{{Y.R. Shimizu, P. Donati, and R.A. Broglia, Phys.
  Rev. Lett. {\bf 85}, 2260 (2000)}}}.

\bibitem[{\citenamefont{{K. Hagino and G.F. Bertsch, Nucl. Phys. {\bf A679},
  163 (2000)}}()}]{[Hag00]}
\bibinfo{author}{\bibnamefont{{K. Hagino and G.F. Bertsch, Nucl. Phys. {\bf
  A679}, 163 (2000)}}}.

\bibitem[{\citenamefont{{W. Nazarewicz, Nucl. Phys. {\bf A557}, 489c
  (1993)}}()}]{[Naz93]}
\bibinfo{author}{\bibnamefont{{W. Nazarewicz, Nucl. Phys. {\bf A557}, 489c
  (1993)}}}.

\bibitem[{\citenamefont{{N. Hinohara {\it et al.}, 
  Phys. Rev. C \textbf{82}, 064313 (2010)}}()}]{[Hin10]}
\bibinfo{author}{\bibnamefont{{N. Hinohara {\it et al.},  Phys. Rev. C \textbf{82}, 064313 (2010)}}}.

\bibitem[{\citenamefont{{J. Terasaki and J. Engel, Phys. Rev. C \textbf{82},
  034326 (2010)}}()}]{[Ter10]}
\bibinfo{author}{\bibnamefont{{J. Terasaki and J. Engel, Phys. Rev. C
  \textbf{82}, 034326 (2010)}}}.

\bibitem[{\citenamefont{{C. Losa {\it et al.}, Phys.
  Rev. C \textbf{81}, 064307 (2010)}}()}]{[Losa10]}
\bibinfo{author}{\bibnamefont{{ C. Losa {\it et al.},
  Phys. Rev. C \textbf{81}, 064307 (2010)}}}.

\bibitem[{\citenamefont{{J. Terasaki and J. Engel, arXiv:1105.3817v1 (2011).}}()}]{[Terasaki:11]}
\bibinfo{author}{\bibnamefont{{J. Terasaki and J. Engel, arXiv:1105.3817v1 (2011).}}}

\bibitem[{\citenamefont{{S. P\'eru  {\it et al.},
  Phys. Rev. C {\bf 83}, 014314 (2011)}}()}]{[Peru.ea:11]}
\bibinfo{author}{\bibnamefont{{S. P\'eru  {\it et
  al.}, Phys. Rev. C {\bf 83}, 014314 (2011)}}}.

\bibitem[{\citenamefont{{M. Martini, S. P\'eru, and M. Dupuis,
  arXiv:1103.1553v1  (2011)}}()}]{[Martini:11]}
\bibinfo{author}{\bibnamefont{{M. Martini, S. P\'eru, and M. Dupuis,
  arXiv:1103.1553v1  (2011)}}}.

\bibitem[{\citenamefont{{J. Terasaki {\it et al.}, Phys.
  Rev. C \textbf{71}, 034310 (2005)}}()}]{[jun05]}
\bibinfo{author}{\bibnamefont{{J. Terasaki {\it et al.},
  Phys. Rev. C \textbf{71}, 034310 (2005)}}}.

\bibitem[{\citenamefont{{T. Nakatsukasa, T. Inakura, and K. Yabana, Phys. Rev.
  C \textbf{76} 024318 (2007)}}()}]{[Nak07]}
\bibinfo{author}{\bibnamefont{{T. Nakatsukasa, T. Inakura, and K. Yabana, Phys.
  Rev. C \textbf{76} 024318 (2007)}}}.

\bibitem[{\citenamefont{{J. Toivanen  {\it et
  al.}, Phys. Rev. C \textbf{81}, 034312 (2010)}}()}]{[Toi10]}
\bibinfo{author}{\bibnamefont{{J. Toivanen {\it
  et al.}, Phys. Rev. C \textbf{81}, 034312 (2010)}}}.

\bibitem[{\citenamefont{{T. Inakura, T. Nakatsukasa, and K. Yabana, Phys. Rev.
  C \textbf{80} 044301 (2009)}}()}]{[Inakura09]}
\bibinfo{author}{\bibnamefont{{T. Inakura, T. Nakatsukasa, and K. Yabana, Phys.
  Rev. C \textbf{80} 044301 (2009)}}}.

\bibitem[{\citenamefont{{T. Inakura, T. Nakatsukasa, and K. Yabana,
  arXiv:1106.3618 (2011)}}()}]{[Inakura11]}
\bibinfo{author}{\bibnamefont{{T. Inakura, T. Nakatsukasa, and K. Yabana,
  arXiv:1106.3618 (2011)}}}.

\bibitem[{\citenamefont{{P. Avogadro and T. Nakatsukasa, Phys. Rev. C in press;
  arXiv:1104.3692.}}()}]{[Avo11]}
\bibinfo{author}{\bibnamefont{{P. Avogadro and T. Nakatsukasa, Phys. Rev. C in
  press; arXiv:1104.3692.}}}

\bibitem[{\citenamefont{{J. Toivanen {\it et al.}, presented at the ARIS-2011
  Conference, May 29-June 3, 2011, Leuven, Belgium; to be
  published}}()}]{ArnoldiQRPA}
\bibinfo{author}{\bibnamefont{{J. Toivanen {\it et al.}, presented at the
  ARIS-2011 Conference, May 29-June 3, 2011, Leuven, Belgium; to be
  published}}}.

\bibitem[{\citenamefont{{C.W. Johnson, G.F. Bertsch, and W.D. Hazelton, Comput.
  Phys. Commun. {\bf 120}, 155 (1999)}}()}]{[Johnson]}
\bibinfo{author}{\bibnamefont{{C.W. Johnson, G.F. Bertsch, and W.D. Hazelton,
  Comput. Phys. Commun. {\bf 120}, 155 (1999)}}}.

\bibitem[{\citenamefont{{C.G. Broyden, Math. Comput. \textbf{19}, 577
  (1965)}}()}]{[broyden]}
\bibinfo{author}{\bibnamefont{{C.G. Broyden, Math. Comput. \textbf{19}, 577
  (1965)}}}.

\bibitem[{\citenamefont{{D.D. Johnson, Phys. Rev. B \textbf{38}, 12807
  (1988)}}()}]{[gbroyden]}
\bibinfo{author}{\bibnamefont{{D.D. Johnson, Phys. Rev. B \textbf{38}, 12807
  (1988)}}}.

\bibitem[{\citenamefont{{A. Baran  {\it et al.}, Phys.
  Rev. C \textbf{78}, 014318 (2008)}}()}]{[hfbbroyden]}
\bibinfo{author}{\bibnamefont{{A. Baran  {\it et al.},
  Phys. Rev. C \textbf{78}, 014318 (2008)}}}.

\bibitem[{\citenamefont{{L. A. Malov and V. G. Soloviev, Nucl. Phys.
  \textbf{A270}, 87 (1976)}}()}]{[Malov76]}
\bibinfo{author}{\bibnamefont{{L. A. Malov and V. G. Soloviev, Nucl. Phys.
  \textbf{A270}, 87 (1976)}}}.

\bibitem[{\citenamefont{{L. A. Malov, V. O. Nestorenko, and V. G. Soloviev, Theor. Math. Phys. {\bf 32}, 646
  (1977)}}()}]{[Malov77]}
\bibinfo{author}{\bibnamefont{{L. A. Malov, V. O. Nestorenko, and V. G.
  Soloviev, Theor. Math. Phys. {\bf 32},
  646 (1977)}}}.

\bibitem[{\citenamefont{{ J. Kvasil {\it et al.}, 
 Phys. Rev. C {\bf 58}, 209 (1998)}}()}]{[Kvasil98]}
\bibinfo{author}{\bibnamefont{{J. Kvasil {\it et al.}, 
 Phys. Rev. C {\bf 58}, 209 (1998)}}}.

\bibitem[{\citenamefont{{C. T.\ Kelley, Iterative Methods for Linear and
  Nonlinear Equations, In {\em SIAM Frontiers in Applied Mathematics}, SIAM,
  Philadelphia, 1995, no. 16}}()}]{[Kelley95]}
\bibinfo{author}{\bibnamefont{{C. T.\ Kelley, Iterative Methods for Linear and
  Nonlinear Equations, In {\em SIAM Frontiers in Applied Mathematics}, SIAM,
  Philadelphia, 1995, no. 16}}}.

\bibitem[{\citenamefont{{E. Chabanat  {\it et al.}, Nucl.
  Phys. {\bf A635}, 231 (1998)}}()}]{sly4}
\bibinfo{author}{\bibnamefont{{E. Chabanat {\it et al.},
  Nucl. Phys. {\bf A635}, 231 (1998)}}}.

\bibitem[{\citenamefont{{M.V. Stoitsov {\it et
  al.}, Comput. Phys. Commun. {\bf 167}, 43 (2005)}}()}]{[Sto05]}
\bibinfo{author}{\bibnamefont{{M.V. Stoitsov  {\it et al.}, Comput. Phys. Commun. {\bf 167}, 43 (2005)}}}.

\bibitem[{\citenamefont{{E. Lipparini and S. Stringari, Phys. Rep. {\bf 175},
  103 (1989)}}()}]{[Lip89]}
\bibinfo{author}{\bibnamefont{{E. Lipparini and S. Stringari, Phys. Rep. {\bf
  175}, 103 (1989)}}}.

\bibitem[{\citenamefont{{U. Garg {\it et al.}, Phys.
  Rev. Lett. \textbf{45}, 1670 (1980)}}()}]{[Garg80]}
\bibinfo{author}{\bibnamefont{{U. Garg  {\it et al.},
  Phys. Rev. Lett. \textbf{45}, 1670 (1980)}}}.

\end{thebibliography}

\end{document}